\RequirePackage{fix-cm}
\documentclass[twocolumn,epjc3]{svjour3}  
\smartqed  
\RequirePackage{graphicx}
\RequirePackage{amsmath}
\RequirePackage[abs]{overpic}
\journalname{Eur. Phys. J. C}
\begin{document}
\title{Exclusive diffractive resonance production in proton-proton collisions at high energies}


\author{R. Fiore\thanksref{e1,addr1}
        \and
        L. Jenkovszky\thanksref{e2,addr2} 
        \and
        R. Schicker\thanksref{e3,addr3} 
}

\thankstext{e1}{e-mail: roberto.fiore@fis.unical.it}
\thankstext{e2}{e-mail: jenk@bitp.kiev.ua}
\thankstext{e3}{e-mail: schicker@physi.uni-heidelberg.de}


\institute{ Department of Physics, University of Calabria, I-87036 Arcavacata di Rende, Cosenza, ITALY \label{addr1}
           \and
Bogolyubov Institute for Theoretical Physics (BITP),
Ukrainian National Academy of Sciences 14-b, Metrologicheskaya
str., Kiev, 03680, UKRAINE \label{addr2}
           \and
Physikalisches Institut, Im Neuenheimer Feld 226, 
Heidelberg University, 69120 Heidelberg, GERMANY \label{addr3}
}

\date{Received: date / Accepted: date}

\maketitle

\begin{abstract}
A model for exclusive diffractive resonance production in proton-proton
collisions at high energies is presented. This model is able to predict double
differential distributions with respect to the mass and the transverse momentum
of the produced resonance in the mass region \mbox{$\sqrt{M^2}\le$5 GeV}.
The model is based on convoluting the Pomeron distribution in the proton
with the Pomeron-Pomeron-meson total cross section. The Pomeron-Pomeron-meson
cross section is saturated by direct-channel contributions from the Pomeron
as well as from two different $f$ trajectories, accompanied by the isolated
f$_0(500)$  resonance dominating the \mbox{$\sqrt{M^{2}}\!\leq\!1\!$ GeV}
region. A slowly varying background is taken into account.


\end{abstract}

\section{Introduction}
\label{Intro}

Central production in proton-proton collisions has been analysed from the low 
energy range $\sqrt{s}$ = 12.7-63 GeV of the  ISR at CERN up to the presently 
highest energy available of $\sqrt{s}$ = 13 TeV achieved in Run II at the LHC.
The CDF Collaboration has analysed central exclusive pion pair production
in proton-antiproton collisions at the TEVATRON energies of
$\sqrt{s}$ =  0.9 and \mbox{1.96 TeV \cite{CDF-cep}.} A physics programme
of \mbox{central exclusive} production in proton-proton collisions
is being pursued by the STAR Collaboration at RHIC \cite{STAR-cep}.
Results from analyses of centrally produced exclusive two track
events from Run I and II at the LHC  are available from the
ALICE \cite{ALICE-cep}, ATLAS \cite{ATLAS-cep}, CMS \cite{CMS-cep} and
LHCb \cite{LHCb-cep} experiments. A comprehensive survey of central exclusive
production has recently been published in a \mbox{review article \cite{Albrow1}.}

The analysis of centrally produced states necessitates the simulation of such
events to study the acceptance and efficiency of the complex large detector
systems. Such events can either be recognized by
identifying rapidity gaps, or by measuring the very forward
scattered protons with Roman Pots. In order to be able to compare the efficiency
of these two approaches, the information of the complete kinematics of the
final state is needed, including the central state and the two outgoing protons.
With the ongoing detector upgrade programmes at RHIC and at the LHC,
much larger data samples of central production events are expected in the
next few years. These larger data samples will allow much improved
data analyses of differential distributions, such as for example
the investigation of resonance parameters by a Partial Wave Analysis (PWA). 
The aim of the study  presented here is the formulation of a model
for simulating such differential distributions.

This article is organized as follows. In the introduction in Sec. 1, the
development of a model for central exclusive production in proton-proton
collisions is motivated. In Sec. 2, the experimental situation
regarding exclusive resonance production at RHIC, the TEVATRON and the LHC
is reviewed. In Sec. 3, the basics of the Regge pole model used for extracting
the Pomeron-Pomeron-meson cross section are summarized.
The total meson production cross section at hadron level is defined in Sec. 4.
The  meson cross section double differential with respect to meson
mass and transverse momentum is derived in Sec. 5.
A summary and an outlook for future studies of the topic presented here
is given in \mbox{Sec. 6.} The parameterisation  of the proton-Pomeron
kinematics is explained in Appendix A. In Appendix B, the
parameterisation of the Pomeron-Pomeron-meson kinematics is
derived. The parameterisation of the three-body final state phase space is
defined in Appendix C.

\section{Central exclusive production at the TEVATRON, at RHIC and at the LHC}
\label{CEPdata}

Central production in proton-proton collisions is characterized 
by the hadronic state produced at or close to mid-rapidity,
and by the two forward scattered protons, or remnants thereof.
No particles are produced between the mid-rapidity system 
and the two beam rapidities on either side of the central system.
Experimentally, these event topologies can be identified  by 
the presence of the two rapidity gaps, by detecting the forward protons 
or its remnants, or by a combination of these two approaches.
Forward scattered neutral fragments can, for example, be measured 
in Zero Degree Calorimeters.

\begin{figure}[h]
\includegraphics[width=0.48\textwidth]{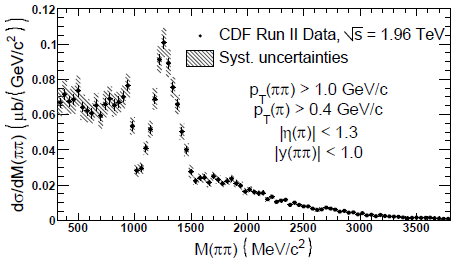}
\caption{Invariant mass distribution of pion pairs measured by the CDF
  Collaboration (Figure taken from Ref. \cite{CDF-cep}).}
\label{fig1}
\end{figure}

The mass distribution of exclusively produced pion pairs shown in
Fig. \ref{fig1} is measured by the CDF Collaboration at the TEVATRON
energy of \mbox{$\sqrt{s}$ = 1.96 TeV \cite{CDF-cep}}. A clear signal
is seen in the region of the f$_{2}$(1270) resonance. The acceptance
of pairs at low masses \mbox{$M<\,$1\,GeV} and low transverse
momenta $p_{T}$ is severely reduced, as shown for example on page 11 of
\mbox{Ref. \cite{CDF-acc}.} A low pair transverse momentum threshold is
hence applied in the data shown in Fig. \ref{fig1}.
For pair masses $M<$ 1 GeV, only the high-end tail of
the tranverse momentum distribution is visible, and conclusions on possible
resonance structures at masses below 1 GeV are therefore difficult to infer from
the data shown in Fig. \ref{fig1}.

\begin{figure}[h]
\includegraphics[width=0.48\textwidth]{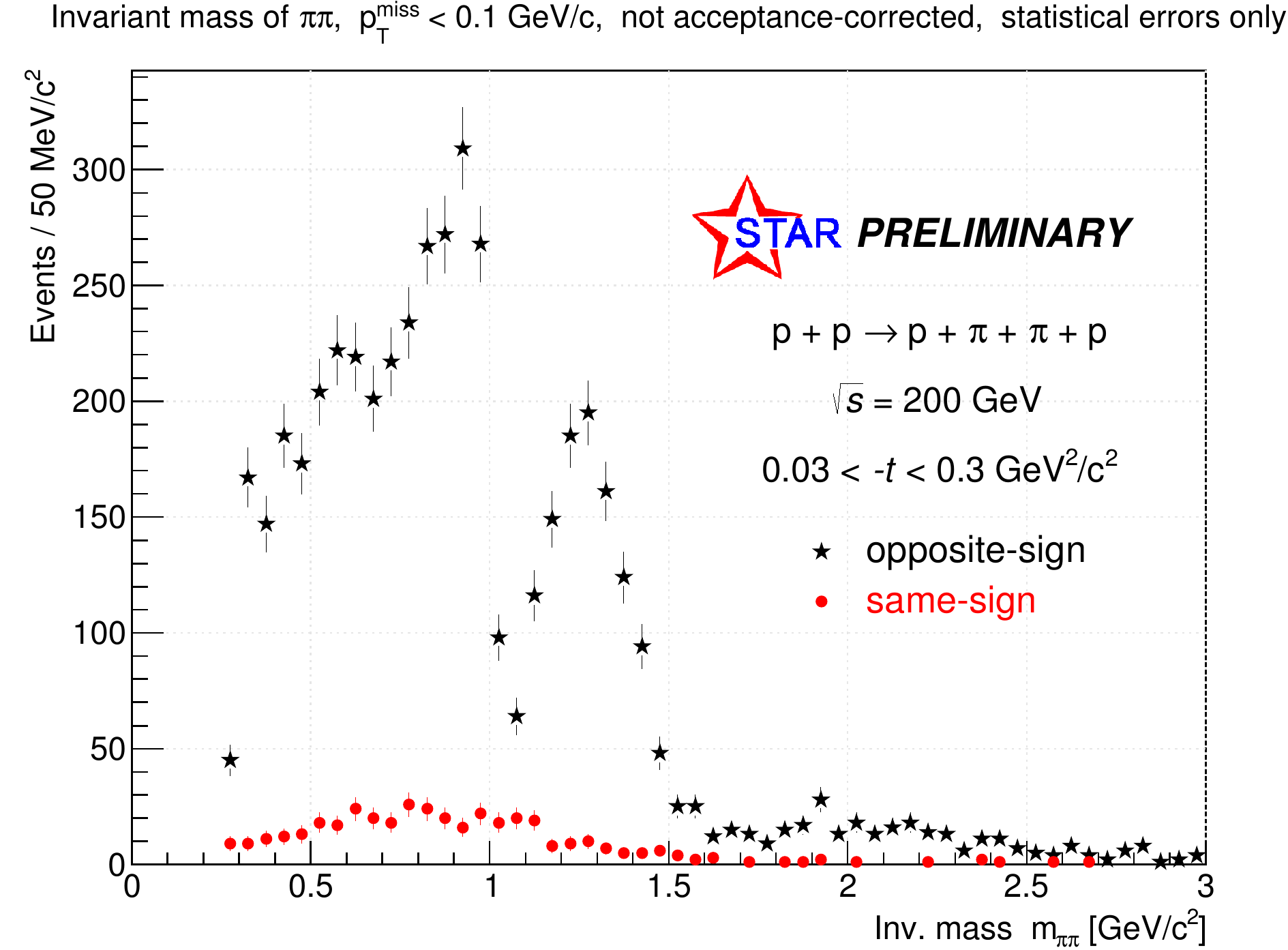}
\caption{Invariant mass distributions of pion pairs measured by the STAR
  Collaboration (Figure taken from Ref. \cite{STAR-cep}).}
\label{fig2}
\end{figure}

The invariant mass distribution of exclusively produced pion pairs measured by
the STAR Collaboration at RHIC at $\sqrt{s}$ = 200 GeV is shown in
Fig. \ref{fig2}. The exclusivity of the event is provided by measuring the pion
pair in the STAR central barrel, and the forward scattered protons in Roman
Pots installed around 16 m away from the interaction point. The opposite-sign
pair distribution is displayed in black, whereas the like-sign pair
distribution reflecting the background is shown in red. Clearly visible is
a strong signal in the mass regions of the f$_{0}$(980) and
the f$_{2}$(1270) resonance.

\begin{figure}[h]
\includegraphics[width=0.48\textwidth]{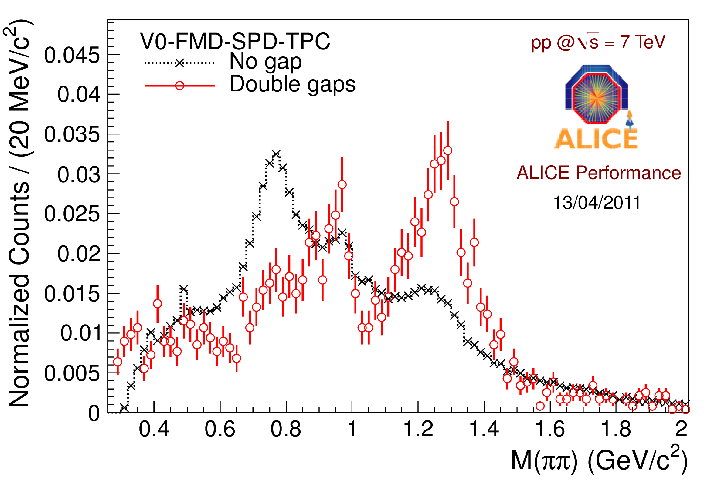}
\caption{Invariant mass distributions of pion pairs measured by the ALICE
  Collaboration (Figure taken from Ref. \cite{ALICE-cep}).}
\label{fig3}
\end{figure}

The invariant mass distributions of pion pairs shown in Fig. \ref{fig3} were
taken by the ALICE Collaboration at mid-rapidity in Run I of the LHC
\cite{ALICE-cep}. This distribution is shown in red for double gap events,
and in black for minimum bias events. The two distributions are normalized
to unity for better comparison of the shape. Clearly visible is an enhancement
of the f$_{0}$(980) and f$_{2}$(1270) resonances in the red distribution, thereby
validating the double gap technique at LHC energies.

\begin{figure}[h]
\includegraphics[width=0.48\textwidth]{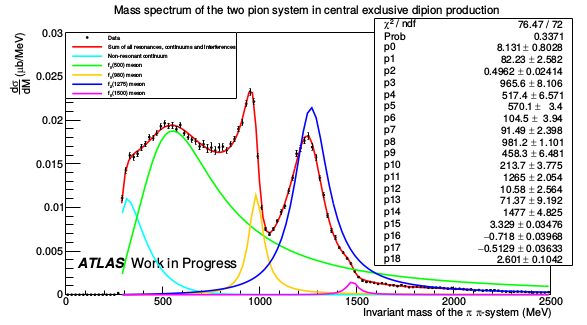}
\caption{The differential cross section measured by the ATLAS Collaboration
  as function of the dipion invariant mass (Figure taken from
  Ref. \cite{ATLAS-cep}).}
\label{fig4}
\end{figure}

In Fig. \ref{fig4}, the central exclusive differential cross section
$pp \rightarrow p + \pi\pi + p$ analysed by the ATLAS Collaboration is
displayed as function of the dipion invariant mass \cite{ATLAS-cep}.
In these events, the pions are identified in the ATLAS central barrel,
and the final state protons are measured in the ALFA detector.
Shown in Fig. \ref{fig4} are  bare Breit-Wigner fits without interference
for the individual resonances (f$_{0}$(980) in yellow, f$_{2}$(1270) in blue),
and the sum of all resonances, continuums and interferences (in red).

\begin{figure}[h]
\includegraphics[width=0.46\textwidth]{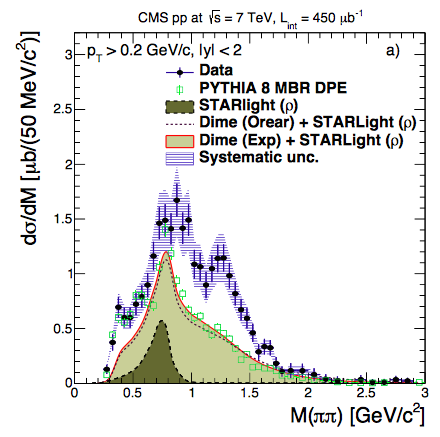}
\caption{Differential cross section measured by the CMS Collaboration as
  function of the pion pair invariant mass (Figure taken from
  Ref. \cite{CMS-cep}).}
\label{fig5}
\end{figure}

In Fig. \ref{fig5}, the differential cross section
$pp \rightarrow p(p^{*}) + (\pi^{+}\pi^{-}) + p (p^{*}) $ measured by
the CMS Collaboration is shown as function
of the pion pair invariant mass, and compared to predictions from
DIME (solid and dashed curves) added to $\rho$ photoproduction from
STARlight (long dashed curve) \cite{CMS-cep}. The results are also compared
to the predictions by PYTHIA and MBR (open squares).

\begin{figure}[h]
\includegraphics[width=0.48\textwidth]{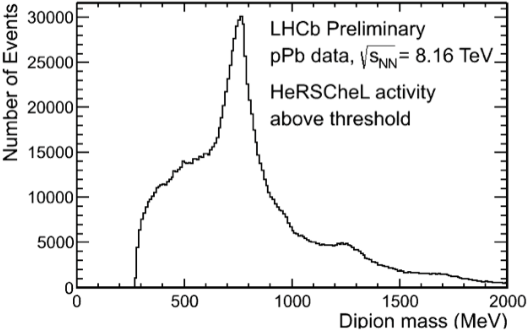}
\caption{Invariant mass distribution of pion pairs measured by the LHCb
  Collaboration (Figure taken from Ref. \cite{LHCb-cep}).}
\label{fig6}
\end{figure}

In Fig. \ref{fig6}, the invariant mass distribution of pion pairs in the
pseudorapidity range 2.0 $< \eta <$ 5.0 taken by the LHCb Collaboration in
proton-lead collisions at $\sqrt {s_{NN}}$ = 8.16 TeV is displayed \cite{LHCb-cep}.
This pair distribution is derived from events where there are exactly
two oppositely charged pions in the event and significant energy in the
forward shower counter HeRSCel, suggesting proton dissociation.
This histogram is dominated by the photoproduction of the $\rho$-resonance,
with a visible signal of the f$_{2}$(1270) resonance, and an indication
of a signal of the f$_{0}$(980).

The comparison of the pion pair mass spectra of the CDF, STAR, ALICE, ATLAS, CMS
and LHCb experiments shown in Figs. \ref{fig1}-\ref{fig6} clearly demonstrates
the existence of the f$_{0}$(980) and the f$_{2}$(1270) resonance in central
exclusive production. The in-depth analysis of these two resonances will play
a pivotal role in future studies of central exclusive production at high
energies. Such analyses provide the opportunity to compare the results from
high statistics data of the different experiments.

\section{Dual resonance model of Pomeron-Pomeron scattering}
\label{PPscat}

We summarize here the main ideas of the dual resonance model of
Pomeron-Pomeron scattering. A detailed discussion of this model
is presented in Ref. \cite{OPUS1}

The triple Reggeon formalism is used in most studies on single and
double diffraction dissociation, and in central diffraction. This approach
is valid in the smooth Regge region, beyond the resonance region,
but is not useful at  low masses which is dominated
by resonances. We solve this  problem by using a dual model.

\begin{figure*}[htb]
\includegraphics[width=.19\textwidth]{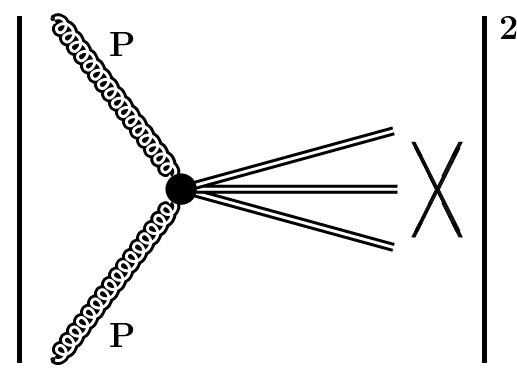}
\includegraphics[width=.038\textwidth]{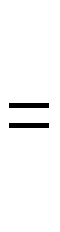}
\hspace{-0.2cm}
\includegraphics[width=.154\textwidth]{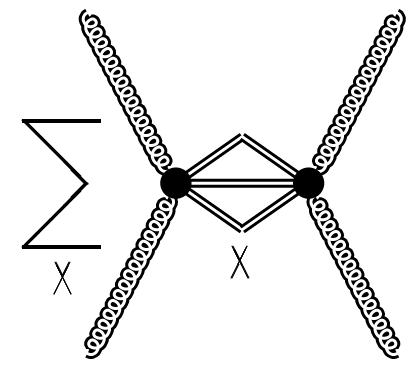}
\hspace{0.2cm}
\includegraphics[width=.037\textwidth]{fig7f.png}
\begin{overpic}[width=.12\textwidth]{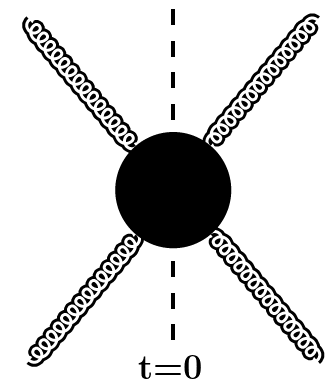}
\put(-34.,20.){\bf Unitarity}
\end{overpic}
\hspace{-0.2cm}
\includegraphics[width=.037\textwidth]{fig7f.png}
\includegraphics[width=.132\textwidth]{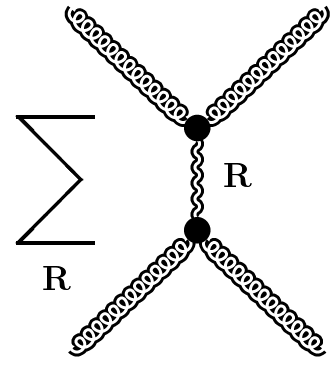}
\hspace{0.2cm}
\includegraphics[width=.038\textwidth]{fig7f.png}
\hspace{0.2cm}
\begin{overpic}[width=.14\textwidth]{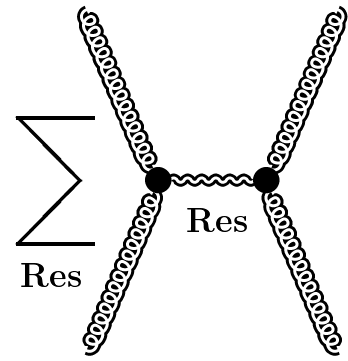}
\put(-50.,20.){\bf Veneziano }
\put(-38.,10.){\bf duality }
\end{overpic}
\caption{Connection, through unitarity (generalized optical
theorem) and Veneziano-duality, between the Pomeron-Pomeron cross section 
and the sum of direct-channel resonances.}
\label{fig7}
\end{figure*}

A sequence of resonances contributes at low masses, where
the one-by-one account of single resonances is possible, but not 
economic for the calculation of cross sections. These resonances
overlap and gradually disappear in the higher mass continuum.
Based on the idea of duality with a limited number of resonances 
lying on non-linear Regge trajectories, an approach to account for many
resonances was suggested in Ref. \cite{JLAB}. 
This approach was used in Ref. \cite{Orava} to calculate low 
mass single- and double-diffractive dissociation at the LHC.  

The main idea behind this approach is illustrated in Fig. \ref{fig7}, 
realized by dual amplitudes with Mandelstam analyticity (DAMA) \cite{DAMA}.
For $s\rightarrow\infty$ and fixed $t$ it is Regge-behaved.
Contrary to the Veneziano model, DAMA not only allows for, 
but rather requires the use of non-linear complex trajectories
providing the resonance widths via the imaginary part of the
trajectory. In the case of limited real part, a finite number 
of resonances is produced. More specifically, the asymptotic 
rise of the trajectories in DAMA is limited by the condition, 
in accordance with an important upper bound, 
\begin{eqnarray}
\big|{\alpha(s)\over{\sqrt s\ln s}}\big|\leq const, \ \
s\rightarrow\infty.
\label{eq1}
\end{eqnarray}

In our study of central production, the direct-channel pole decomposition 
of the dual amplitude $A(M_{X}^{2},t)$ is relevant. Different trajectories 
$\alpha_{i}(M_X^2)$ \mbox{contribute to the} amplitude, with $\alpha_{i}(M_X^2)$
a  non-linear, complex Regge trajectory in the Pomeron-Pomeron (PP) system,

\begin{equation}
A(M_X^2,t)=a\sum_{i=f,P}\sum_{J}\frac{[f_{i}(t)]^{J+2}}{J-\alpha_i(M_X^2)}.
\label{eq2}
\end{equation}

In Eq. (\ref{eq2}), the pole decomposition of the dual amplitude 
$A(M_{X}^{2},t)$ is shown with $t$ the squared momentum transfer in the 
$PP\rightarrow PP$ reaction. The index $i$ sums over the trajectories which 
contribute to this amplitude. Within each trajectory, the second sum
extends  over the bound states of spin $J$. The prefactor $a$
in \mbox{Eq. (\ref{eq2})} is  of numerical value a = 1 GeV$^{-2}$ = 0.389 mb.

The pole residue $f(t)$ appearing in the $PP\rightarrow PP$ system
is fixed by the dual model, in particular by the compatibility of its Regge
asymptotics with Bjorken scaling and reads

\begin{equation}
\label{eq3} f(t) = (1-t/t_0)^{-2},
\end{equation}

where $t_0$ is a parameter to be fitted to the data. However, due to
the absence of data so far, we set $t_0=0.71$ GeV$^2$ for the moment
as in the proton elastic form factor. Note that the residue enters with a
power ($J\!+\!2$) in \mbox{Eq. (\ref{eq2}),} thereby strongly damping 
higher spin resonance contributions. The imaginary part of the 
amplitude $A(M_X^2,t)$ given in Eq. (\ref{eq2}) is defined by 
\begin{equation*}
\Im m\, A(M_{X}^2,t)=
\end{equation*}
\begin{equation}
a\sum_{i=f,P}\sum_{J}\frac{[f_{i}(t)]^{J+2} 
\Im m\,\alpha_{i}(M_{X}^2)}{(J-Re\,\alpha_{i}(M_{X}^2))^2+ 
(\Im m\,\alpha_{i}(M_{X}^2))^2}.
\label{eq4}
\end{equation}

For the $PP$ total cross section we use the norm 
\begin{equation}
\sigma_{t}^{PP} (M_{X}^2)= {\Im m\; A}(M_{X}^2, t=0),
\label{eq:ppcross}
\end{equation}

and recall that the amplitude $A$ and the cross section $\sigma_{t}$ carry 
dimensions of mb due to the parameter $a$ discussed above.
The Pomeron-Pomeron channel, $PP\rightarrow M_X^2$, couples to the Pomeron
and $f$ channels dictated by conservation of quantum numbers.  For calculating
the $PP$ cross section, we hence consider the trajectories associated
to the f$_0$(980) and f$_2$(1270) resonance which are seen as strong
signals in the experiment data as shown in Sec. \ref{CEPdata},
and the Pomeron trajectory \cite{OPUS1}. 

The experimental data on central exclusive pion pair production measured at 
the ISR, RHIC, TEVATRON and the LHC all show a broad 
continuum for pair masses $m_{\pi^+\pi^-}<$ 1 GeV. This mass region is 
experimentally difficult to access due to the missing acceptance for 
pairs of low mass and low transverse \mbox{momentum $p_{T}$ \cite{CDF-acc}.}
The population of this  mass region is attributed to the f$_{0}$(500), a 
resonance which has been controversial for many decades.
The non-ordinary nature of the f$_{0}$(500) state is  
corroborated by the fact that it does not fit into the Regge description of 
classifying $q\bar{q}$-states into trajectories \cite{Anisovich}. 
In spite of the complexity of the f$_{0}$(500) resonance, and the controversy 
on its interpretion and description, we take the practical but 
simple-minded approach of a Breit-Wigner resonance \cite{PDG} 

\begin{eqnarray} 
A(M^{2}) = a\; \frac{-M_0\Gamma}{M^{2}-M_{0}^{2}+iM_{0}\Gamma}. 
\label{eq:BWampl}
\end{eqnarray} 

In Eq. (\ref{eq:BWampl}), the parameterisation of the relativistic Breit-Wigner 
amplitude is shown with $M_{0}$ and $\Gamma$ the mass and width, respectively. 
The Breit-Wigner amplitude of Eq. (\ref{eq:BWampl}) is used below  for 
calculating the contribution of the f$_{0}$(500) resonance to the 
PP cross section.

The PP cross section is calculated from the 
imaginary part of the amplitude by use of the optical theorem
\begin{equation*}
\sigma_{t}^{PP} (M^2) \;\; = \;\; {\Im m\; A}(M^2, t=0) \;\; = 
\end{equation*}
\begin{equation}
\;\; a\sum_{i=f,P}\sum_{J}\frac{[f_{i}(0)]^{J+2}\; \Im m \;\alpha_{i}(M^{2})}
{(J-\Re e \;\alpha_{i}(M^{2}))^{2}+(\Im m \;\alpha_{i}(M^{2}))^{2}}.
\label{eq:imampl}
\end{equation}

In Eq. (\ref{eq:imampl}), the index $i$ sums over the trajectories which
contribute to the cross section, in our case the $f_{1}$, $f_{2}$
and the Pomeron trajectory. For a discussion of the
$f_{1},f_{2}$ trajectories, and a listing of the resonances
defining these trajectories, see Ref. \cite{OPUS1}.
Within each trajectory, the summation extends over the poles
of spin $J$ as expressed by the second summation sign.
The value $f_{i}(0) =f_{i}(t)\big |_{\text t=0}$ is not known a priori,
but can be extracted from the experimental data by analysing 
relative strengths of resonances within a trajectory.
 
The Breit-Wigner parameterisation of the isolated f$_{0}$(500) 
resonance contributes to the cross section with 
\begin{equation} 
\sigma_{\text{f}_{0}(500)}^{PP}(M^{2}) = a \sqrt{1.\!-\!\frac{4\,m_{\pi}^{2}}{M^{2}}}
\frac{{M}_{0}^{2}\Gamma^{2}}{({M}^{2}\!-\!{M}_{0}^{2})^{2}\!+\!{M}_{0}^{2}\Gamma^{2}}, 
\label{eq:BWcross}
\end{equation} 

with the resonance mass of $M_{0}$ = (0.40--0.55) GeV and a width 
$\Gamma$ = (0.40--0.70) GeV \cite{PDG}.  The quantity
$\sqrt{1.-4\,m_{\pi}^{2}/M^{2}}$ in Eq. (\ref{eq:BWcross}) is the 
threshold phase space factor for the two-pion decay.

A background  term of form
\begin{equation}
\sigma_{backgr.}^{PP}(M^{2}) = c\cdot (0.1+ \text{log}(M^{2}))\text{\;mb}
\label{eq:back}
\end{equation}
is added, with the numerical value of the
parameter c fitted to data \cite{Kononenko}.

\begin{figure}[h]
\includegraphics[width=.52\textwidth]{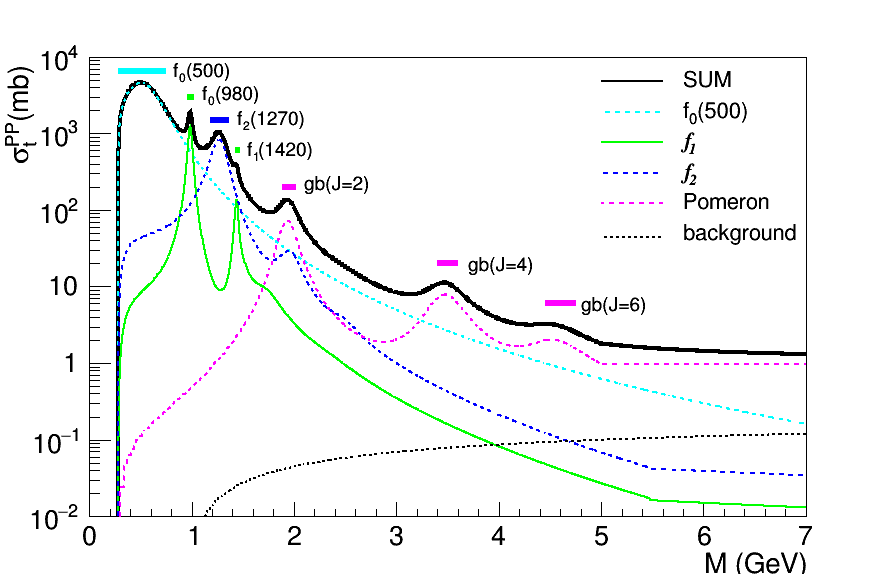}
\caption{Contributions of the f$_{0}$(500) resonance, the $f_{1}$, $f_{2}$ and 
the Pomeron trajectory, and of the background to the PP total cross section.}
\label{fig8}
\end{figure}

The different contributions to the PP total cross section are shown in
Fig. \ref{fig8}. The contribution of the f$_{0}$(500) resonance according
to Eq. (\ref{eq:BWcross}) is displayed by the dashed cyan line, with a
central value for the mass $M_{0}$ as well as for the width $\Gamma$,  
$M_{0} = 475\,$MeV and $\Gamma = 550\,$MeV, respectively.
The contribution of the $f_{1}$ and $f_{2}$ trajectory are shown by the solid
green and by the dashed blue line, respectively. The contribution from the
Pomeron trajectory is displayed by the dashed magenta line. The
Pomeron-Pomeron total cross section is calculated by summing over the
contributions discussed above, and is shown by the solid black line.

\section{Total meson cross section at hadron level}

In order to derive the meson production cross section at hadron level,
we start from the definition of an infinitesimal cross section element

\begin{equation}
d\sigma = \frac{|\mathcal{M}|^{2}}{\text{flux}} dQ,
\label{eq:had1}
\end{equation}

with $\mathcal{M}$ the invariant amplitude, $dQ$ the  Lorentz-invariant
phase space and flux the flux factor. Above Eq. (\ref{eq:had1}) can be
written at hadron and Pomeron level, and hence yields an equation
\begin{equation}
  |\mathcal{M}|^{2} dQ = \text{flux}_{\text{prot}}\; d\sigma_{\text{prot}} =
  \text{flux}_{\text{Pom}} F^{\text{Pom}}_{\text{prot}}\; d\sigma_{\text{Pom}}.
\label{eq:had2}
\end{equation}
In Eq. (\ref{eq:had2}), the parameter $F^{\text{Pom}}_{\text{prot}}$ characterizes
the distribution of Pomerons in the proton. Equation  (\ref{eq:had2}) can be
rearranged, and the infinitesimal cross section element at hadron level
can be expressed as function of the corresponding element at Pomeron level,
\begin{equation}
  d\sigma_{\text{prot}} = \frac{\text{flux}_{\text{Pom}}}{\text{flux}_{\text{prot}}} F^{\text{Pom}}_{\text{prot}} d\sigma_{\text{Pom}}.
\label{eq:had3}
\end{equation}

The flux factor for two-body collisions of A and B
is defined in manifestly invariant form \cite{HalzenMartin},
\begin{equation}
 \text{flux} = 4.*\big( (p_{A}\cdot p_{B})^{2}-m_{A}^{2} m_{B}^{2}\big)^{1/2}. 
  \label{eq:had4}
\end{equation}

The distribution $F^{\text{Pom}}_{\text{prot}}$ of Pomerons in the proton has been
examined in the study of the structure of the Pomeron, and has been
parameterised as function of the two kinematical
variables $(t,\xi)$ \cite{DLpomdist},
\begin{equation}
  F^{\text{Pom}}_{\text{prot}}(t,\xi) =
  \frac{9 \beta_{0}^{2}}{4\pi^{2}}\;\big[F_{1}(t)\big]^{2}\xi^{1-2\alpha(t)}.
\label{eq:had5}
\end{equation}

The distribution $F^{\text{Pom}}_{\text{prot}}$ in Eq. (\ref{eq:had5})
is integrated over the azimuthal angle of the final state proton.
The variable $t$ represents the squared four-momentum transfer of the proton
(Mandelstam $t$), $\xi$ denotes the proton fractional longitudinal momentum
loss, and $\beta_{0}$=1.8 GeV$^{-1}$.  The Regge factor  $\xi^{1-2\alpha(t)}$
enters instead of propagators for the Pomerons, with $\alpha(t)$ the Pomeron
trajectory

\begin{equation}
  \alpha(t)\!=\!1.\!+\!\varepsilon\!+\!\alpha^{'}t, \;\;\;\;\varepsilon \sim 0.085, \;\;\;\;\alpha^{'}\!=\!0.25\: \text{GeV}^{-2}.
\label{eq:had6}
\end{equation}

The factor  $F_{1}(t)$ in Eq. (\ref{eq:had5}) represents the elastic form
factor, and is taken as

\begin{equation}
  F_{1}(t) = \frac{4 m^{2} - 2.8 t}{4 m^{2} - t}\Big(\frac{1}{1-t/0.7}\Big)^{2}. 
\label{eq:had7}
\end{equation}

Note that the form factor as given by Eq. (\ref{eq:had7}) resembles the
pole residue in the dual amplitude as depicted in Eq. (\ref{eq3}),
except for the prefactor which depends on the mass of the proton.

The meson production cross section at hadron level can be expressed as an
integral over the 6 kinematical variables
($t_{A},\xi_{A},\phi_{A},t_{B},\xi_{B},\phi_{B}$)
which parameterise the Pomeron distribution in the two protons,

\begin{equation*}
\sigma_{pp}\!= \!\!\!\int\!\!\!\!\!\int\!\!\!\!\!\int\!\!\!\!\!\int\!\!\!\!\!\int\!\!\!\!\!\int\!\! 
\frac{\text{flux}_{\text{Pom}}}{\text{flux}_{\text{prot}}} \cdot
F^{\text{Pom}}_{\text{prot}_{A}}(t_{A},\xi_{A},\phi_{A})
F^{\text{Pom}}_{\text{prot}_{B}}(t_{B},\xi_{B},\phi_{B})
\end{equation*}
\begin{equation}
\hspace{1.2cm} \times\: \sigma^{PP}(M_{x},t_{A,B})\:dt_{A}d\xi_{A}d
\phi_{A}dt_{B}d\xi_{B}d\phi_{B}.
\label{eq:had8}
\end{equation}

\section{Double differential meson cross section}

From the expression given in Eq. (\ref{eq:had8}), the cross section can be
derived double differentially with respect to the mass $M$ and transverse
momentum $p_{T}$ of the produced meson. Such a derivation can be achieved by a
transformation of the total cross section to the parameters $M$ and $p_{T}$
as integration variables. The Pomeron distribution in the proton as expressed in
Eq. (\ref{eq:had5}) is integrated over the azimuthal angle, and the Pomeron
distributions entering Eq. (\ref{eq:had8}) therefore have no explicit azimuthal
dependence. One of the azimuthal angles of the two protons can hence be
integrated out, and one azimuthal angle $\phi = \phi_{A} - \phi_{B}$
representing the azimuthal opening angle between the two protons remains.

We hence look for a change of variables 
\begin{equation}
(t_{A},\xi_{A},t_{B},\xi_{B},cos(\phi)) \longmapsto (u_{+},u_{-},v_{-}, M, p_{T}). 
\label{eq:had9}
\end{equation}

Such a transformation is accompanied by a transformation of the differentials
according to the Jacobian determinant $J$
\begin{equation}
  dt_{A}\:d\xi_{A}\:dt_{B}\:d\xi_{B}\:dcos(\phi) =
  J\; du_{+}du_{-}dv_{-}dMdp_{T}.
\label{eq:had10}
\end{equation}

This transformation is derived in Appendices A, B and C, with the Jacobian $J$
defined by 
\begin{equation}
  J = \frac{4}{\beta^{2}}
  \frac{2 M p_{T}}{G \gamma^{2} \frac{\partial H}{\partial v_{+}}}.
  \label{eq:had11}
\end{equation}

Here, $G$ and $H$ are functions of the transformed kinematic variables
$(u_{+},u_{-},v_{-}, M, p_{T})$ and are defined in \mbox{Appendix B.}
In these transformed variables, the meson production cross section is written as

\begin{equation*}
\sigma_{pp} =
\!\!\int\!\!\!\!\!\int\!\!\!\!\!\int\!\!\!\!\!\int\!\!\!\!\!\int\!
\frac{\text{flux}_{\text{Pom}}}{\text{flux}_{\text{prot}}} \cdot
\tilde{F}^{\text{Pom}}_{\text{prot}_{A}}
\tilde{F}^{\text{Pom}}_{\text{prot}_{B}}
\sigma^{PP}(M,u_{+},u_{-})
\end{equation*}
\begin{equation}
\hspace{3.2cm}\times\;J\:du_{+}du_{-}dv_{-}dMdp_{T}.
  \label{eq:had12}
\end{equation}

Here, the quantity $\tilde{F}^{\text{Pom}}_{\text{prot}}$ denotes the Pomeron
distribution $F^{\text{Pom}}_{\text{prot}}$ expressed in the transformed variables
$(u_{+},u_{-},v_{-}, M, p_{T})$. The double differential meson cross section can
be extracted from Eq.(\ref{eq:had12}) as

\begin{equation*}
\frac{d\sigma_{pp}}{dMdp_{T}} =
\!\!\!\!\int\!\!\!\!\!\int\!\!\!\!\!\int
\frac{\text{flux}_{\text{Pom}}}{\text{flux}_{\text{prot}}} \cdot
\tilde{F}^{\text{Pom}}_{\text{prot}_{A}}
\tilde{F}^{\text{Pom}}_{\text{prot}_{B}}
\sigma^{PP}(M,u_{+},u_{-})
\end{equation*}
\begin{equation}
\hspace{4.cm}\times\;J\:du_{+}du_{-}dv_{-}.
  \label{eq:had13}
\end{equation}

The numerical evaluation of the double differential cross section
$d\sigma_{pp}/dMdp_{T}$ involves an integration over the three
parameters ($u_{+},u_{-},v_{-}$). The integration limits in this
parameter space can be determined according to the kinematically allowed
range as shown in Fig. \ref{figa2} of Appendix A.
\begin{figure}[b]
  \begin{overpic}[width=.46\textwidth]{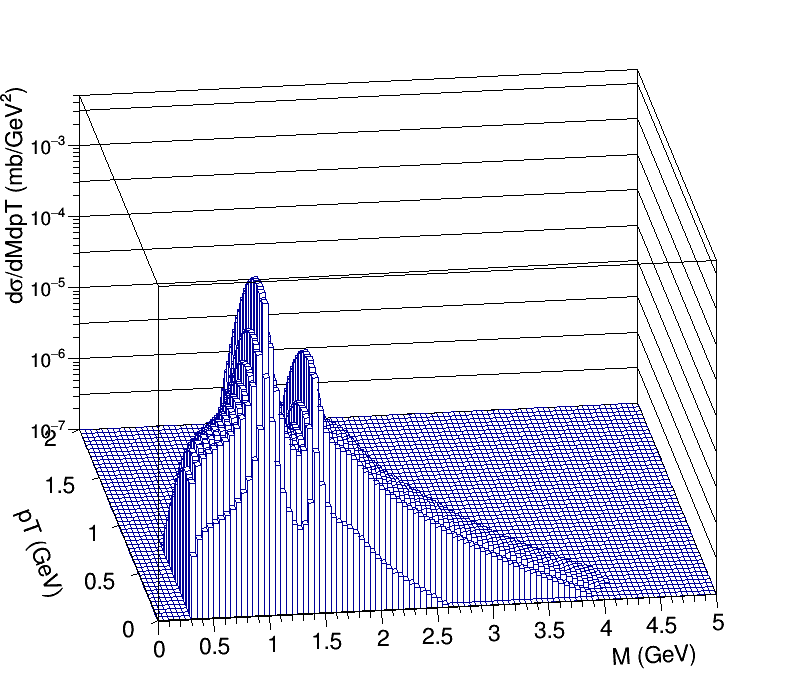}
\end{overpic}
  \caption{Double differential meson cross section for contribution
  of $f_{1}$ trajectory.}
\label{fig9}
\end{figure}

The double differential meson production cross section
$d\sigma_{pp}/dMdp_{T}$ at hadron level is shown in Fig. \ref{fig9}
for the contribution of the $f_{1}$ trajectory.

\begin{figure}[h]
  \begin{overpic}[width=.46\textwidth]{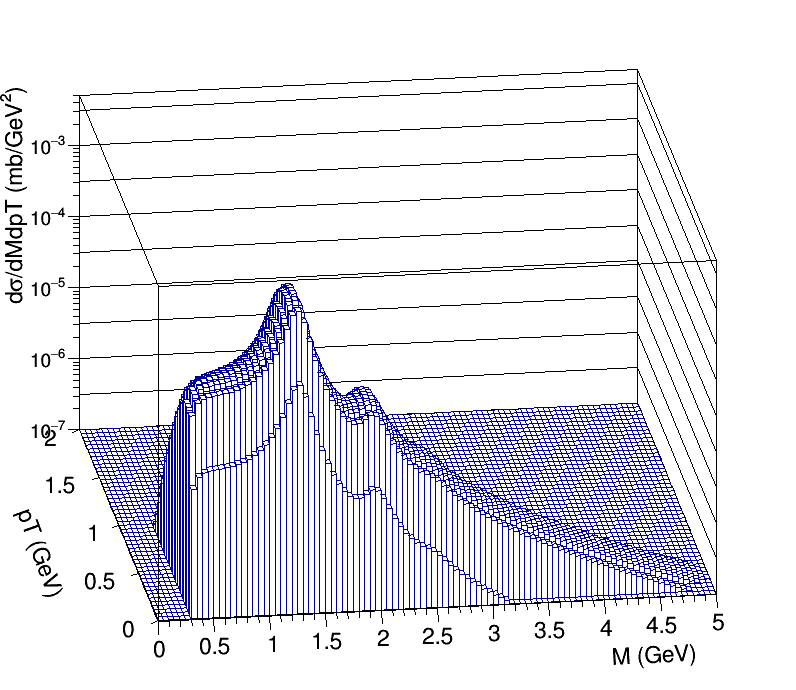}
\end{overpic}
  \caption{Double differential meson cross section for contribution
  of $f_{2}$ trajectory.}
\label{fig10}
\end{figure}

The double differential meson production cross section
$d\sigma_{pp}/dMdp_{T}$ at hadron level is shown in Fig. \ref{fig10}
for the contribution of the $f_{2}$ trajectory.

The model presented here does not make any prediction on the
absolute value of these two contributions to the total meson
production cross section. The strength of these contributions needs to
be extracted from experimental data.

\begin{figure}[h]
\includegraphics[width=.46\textwidth]{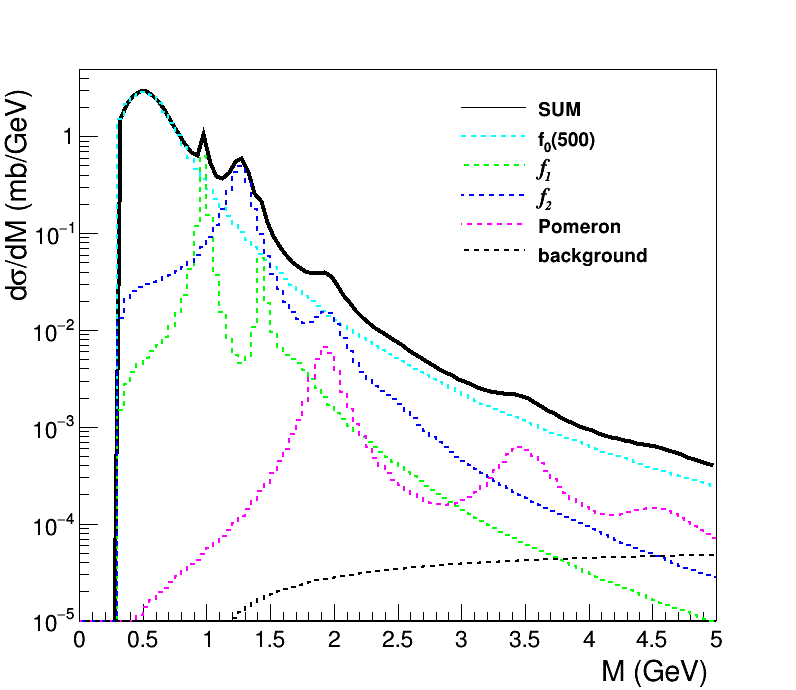}
\caption{Contributions of the f$_{0}$(500) resonance, the $f_{1}$, $f_{2}$ 
  and the Pomeron trajectory, and of the background to the meson
  differential cross section d$\sigma$/dM at hadron level.}
\label{fig11}
\end{figure}

The contributions to the differential meson cross section $d\sigma_{pp}/dM$
at hadron level are shown in Fig. \ref{fig11}. These contributions are derived
by integrating the respective double differential cross section
$d\sigma_{pp}/dMdp_{T}$ over the transverse momentum $p_{T}$. The contribution
of the f$_{0}$(500) resonance results from the Breit-Wigner parameterisation of
Eq. (\ref{eq:BWcross}), and is shown in Fig. \ref{fig11} by the dashed cyan
line. The contributions of the $f_{1}$ and $f_{2}$ trajectory of Figs. \ref{fig9}
and \ref{fig10} are represented by the dashed green and dashed blue line,
respectively. The contribution of the Pomeron trajectory is shown by the dashed
magenta line, whereas the background is represented by the dashed black line. 
The contributions of the f$_{0}$(500), the $f_{1},f_{2}$ and the
Pomeron trajectory, and background, are normalized such that they correspond
to cross sections of 1 mb, 50 $\mu$b, \mbox{120 $\mu$b}, \mbox{2 $\mu$b}
and \mbox{0.1 $\mu$b} per unit of rapidity at mid-rapidity, respectively.

\section{Summary and outlook}

A model is presented  for calculating double differential distributions in
mass and transverse momentum for exclusive diffractive meson production at
high energies. This model is based on convoluting the Pomeron
distribution in the proton with the Pomeron-Pomeron-meson cross section.
Such double differential distributions are presented
for the f$_{0}$(980) and f$_{2}$(1270) resonances.
The absolute contribution of these resonances to the total meson
cross section cannot be derived within this approach, and must hence be
deduced from experimental data.
This model can be expanded to include photon-Pomeron interactions, and
hence distributions of diffractive photoproduction of low mass vector mesons
$\rho,\omega,\phi$ can be calculated. With such a model, a multitude of
resonances with different spin-parity assignments can be generated
in order to validate the fitting procedures of a Partial Wave Analysis
framework. The model presented here can be extended to lower beam energies
where not only Pomeron-Pomeron,  but also Pomeron-Reggeon and Reggeon-Reggeon
diagrams need to be considered. To make realistic predictions for
exclusive resonance production at the LHC,
diffractive excitations of protons must be included.
The results discussed here are necessary and essential for such 
calculations.

\section{Acknowledgements}

This work is supported by the German Federal Ministry of Education and
Research under  promotional reference 05P15VHCA1. One of us (L.J.)
gratefully acknowledges an EMMI visiting Professorship at the University of
Heidelberg where this study was initiated.

\newpage

\section*{Appendix A}

\section*{Kinematics Proton-Pomeron vertex}

The kinematics at the proton-Pomeron vertex is parameterised as function of 
three variables, the squared four-momentum transfer $t$ to the proton 
\mbox{(Mandelstam $t$)}, the fractional longitudinal momentum loss $\xi$
of the proton  $(\xi > 0)$, and the azimuthal angle $\phi$ of the
scattered proton  ($0 \le \phi < 2 \pi $).

\begin{figure}[h]
\begin{center}
\includegraphics[width=.42\textwidth]{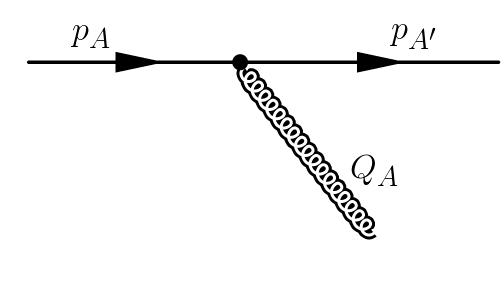}
\end{center}
\caption{Proton-Pomeron vertex.}
\label{figa1}
\end{figure}
The elementary proton-Pomeron vertex is shown in Fig. \ref{figa1}. Here, the 
four-momentum of the proton before and after scattering is denoted by $p_{A}$ 
and $p_{A'}$,  respectively, with $Q_{A}$  the four-momentum of the exchange,
\begin{equation}
Q_{A} = P_{A} - P_{A'}.
\label{eq:a1}
\end{equation}

The squared four-momentum transfer $t$ to the proton is negative ($ t < 0$),
and is  defined by   

\begin{equation}
  t = Q_{A}^{2} = (P_{A} - P_{A'})^{2}.
 \label{eq:a2}
\end{equation}

The components of the four-momentum $P_{A}$ are defined by 
$(E_{A},p_{A,x},p_{A,y},p_{A,z})$, with an analogous definition for $P_{A'}$.
We introduce a temporary variable $\zeta$ for the fractional energy loss 
of the proton $(\zeta > 0)$,
\begin{equation}
\zeta = \frac{E_{A}-E_{A'}}{E_{A}},
\label{eq:a3}
\end{equation}

which can be rewritten for $E_{A'}$ as
\begin{equation}
E_{A'} = E_{A}(1 - \zeta).
\label{eq:a4}
\end{equation}
 
With $\gamma$ the Lorentz-factor of the beam, the four-momenta $P_{A}$, 
$P_{A'}$ and $Q_{A}$ are written as    
\begin{equation}
P_{A} = (\gamma m, \:0, \:0, \:\gamma \beta m),
\label{eq:a5}
\end{equation}
\begin{equation}
P_{A'} = (\gamma m (1\!-\!\zeta), \:p_{T}\:cos \phi, \:p_{T}\:sin\phi, 
\:\gamma \beta m (1\!-\!\xi)),
\label{eq:a6}
\end{equation}
\begin{equation}
Q_{A} = (\gamma m \zeta, \:-p_{T}\:cos\phi, \:-p_{T}\:sin\phi, 
\:\gamma \beta m \xi),
\label{eq:a7}
\end{equation}

with $m$ the proton mass, and $p_{T}$ the transverse momentum of the final state
proton. From Eqs. (\ref{eq:a2}),(\ref{eq:a5}) and (\ref{eq:a6}),
the fractional energy loss $\zeta$ and transverse momentum $p_{T}$ of the
scattered proton are deduced as

\begin{equation}
\zeta = \frac{t}{2\gamma^{2}m^{2}} + \beta^{2}\xi,
\label{eq:a8}
\end{equation}
\begin{equation}
p_{T}^{2} = f(t,\xi) = \frac{t^{2}}{4 \gamma^{2} m^{2}} - t + t \xi \beta^{2} - m^{2} \beta^{2} \xi^{2}.
\label{eq:a9}
\end{equation}
  
The function $f(t,\xi)$ in Eq. (\ref{eq:a9}) can be transformed into a bilinear
form  by quadratic completion. This procedure defines a transformation
to a set of variables on the principal axes of the bilinear form.
These principal axes coordinates $(u,v)$ are determined as    

\begin{equation}
u = \frac{-t}{2 m} +m,
\label{eq:a10}
\end{equation}
\begin{equation}
v = m \beta \xi - \frac{t \beta}{2 m}.
\label{eq:a11}
\end{equation}

The transformation to the variables $(u,v)$ is associated with a
Jacobian determinant  

\begin{equation}
dudv = \Bigl| \frac{\partial(u,v)}{\partial(t,\xi)} \Bigr|\; dtd\xi 
= \frac{\beta}{2}\; dtd\xi.
\label{eq:a12}
\end{equation}

The transformation defined by Eqs. (\ref{eq:a10}),(\ref{eq:a11})
is non-singular due to the constant finite value of the 
Jaco- bian $J$ = $\beta$/2, and can hence be inverted to express
the variables $(t,\xi)$ as function of the new coordinates $(u,v)$ 

\begin{equation}
t = -2m (u-m), 
\label{eq:a13}
\end{equation}
\begin{equation}
\xi = \frac{v}{m \beta} - \frac{u}{m} +1.
\label{eq:a14}
\end{equation}

Our region of interest ($\xi>0, t<0$) is transformed by
Eqs. (\ref{eq:a10}),(\ref{eq:a11}) into a section of the $(u,v)$-plane
defined by ($u > 0, v > 0$). The fractional energy loss $\zeta$ and the
transverse momentum $p_{T}$  are expressed in the new coordinates $(u,v)$ as

\begin{equation}
\zeta = \frac{v \beta}{m} - \frac{u}{m} +1,
\label{eq:a15}
\end{equation}
\begin{equation}
p_{T}^{2} = u^{2} - v^{2} - m^{2}.
\label{eq:a16}
\end{equation}

The loci of constant transverse momentum $p_{T}$ lie on the branch ($u>0, v>0$)
of the hyperbola defined in canonical form by

\begin{equation}
\frac{u^{2}}{p_{T}^{2}+m^{2}} - \frac{v^{2}}{p_{T}^{2}+m^{2}} = 1.
\label{eq:a17}
\end{equation}

\begin{figure}[h]
\begin{center}
\includegraphics[width=.5\textwidth]{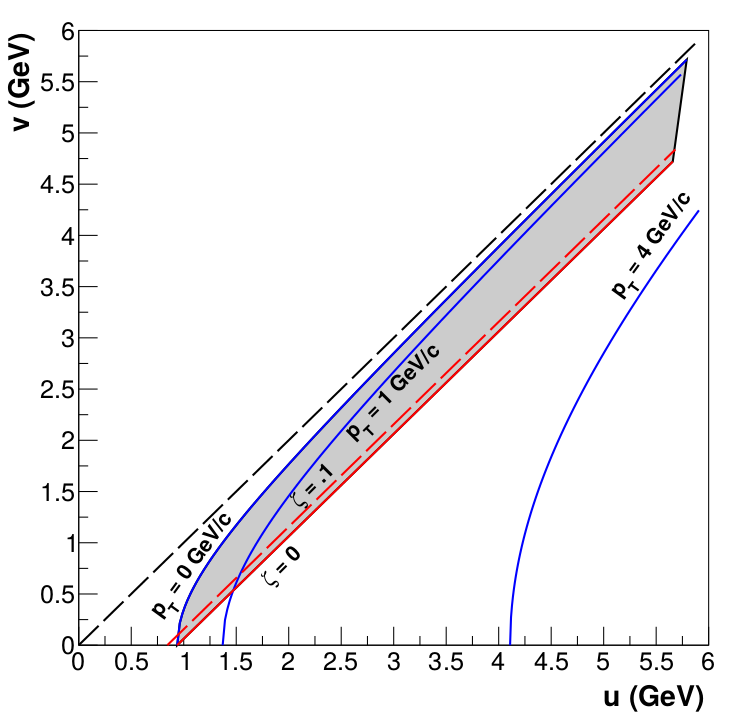}
\end{center}
\caption{Proton-Pomeron kinematics in (u,v)-plane.}
\label{figa2}
\end{figure}

The hyperbolae of transverse momenta $p_{T}$ = 0, 1 and 4 GeV/c  are
shown in Fig. \ref{figa2} by solid blue lines.
The solid and dashed red lines represent the constant fractional energy loss
of  $\zeta = 0$ and $\zeta = 0.1$, respectively. The kinematically allowed
(u,v)-values are indicated by the grey shaded area. The requirement of
positive fractional momentum loss $(\xi >0)$ is  not shown in
\mbox{Fig. \ref{figa2}} since the condition of positive fractional energy
loss $(\zeta > 0)$ is more restrictive. A fixed value of $t=t_{0}$ corresponds
to a value of \mbox{$u_{0}$ =  $m-t_{0}/2m$.} 
The allowed v-values are hence in the range 
defined by the line of energy loss $\zeta=0$ and the 
hyperbola \mbox{$p_{T}$ = 0 GeV/c,}

\begin{equation}
\frac{u_{0}-m}{\beta} < v < \sqrt{u_{0}^{2}-m^{2}}.
\label{eq:a18}
\end{equation}

\section*{Appendix B}

\section*{Kinematics Pomeron-Pomeron-Meson vertex}

The kinematics of the Pomeron-Pomeron-meson vertex is derived from 
two uncorrelated proton-Pomeron vertices.
\begin{figure}[h]
\begin{center}
\includegraphics[width=.36\textwidth]{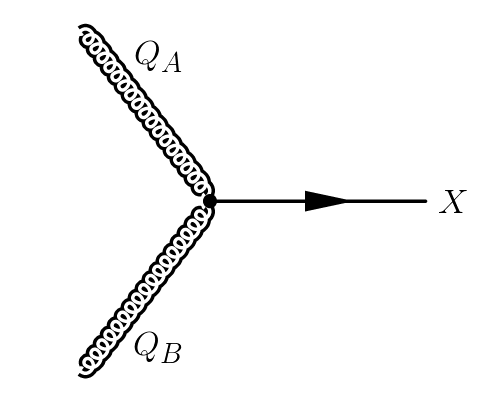}
\end{center}
\caption{Pomeron-Pomeron-meson vertex.}
\label{figb1}
\end{figure}
The exchanges $Q_{A}$ and $Q_{B}$ shown in Fig. \ref{figb1}
combine to generate the meson of four-momentum $X$, 

\begin{equation*}
X = Q_{A} + Q_{B}
\end{equation*}
\begin{equation*}
= (\gamma m (\zeta_{A}+\zeta_{B}), \:-p_{T,A}\:cos\phi_{A}-p_{T,B}\:cos\phi_{B}, 
\end{equation*}
\begin{equation}
\hspace{1.cm}\:-p_{T,A}\:sin\phi_{A}-p_{T,B}\:sin\phi_{B}, \:\gamma \beta m (\xi_{A}-\xi_{B})). 
\label{eq:b1}
\end{equation}

From the four-momentum X shown in Eq. (\ref{eq:b1}), the invariant
mass $M$ of the meson can be derived as

\begin{equation*}
M^{2} = \gamma^{2}m^{2}(\zeta_{A}+\zeta_{B})^{2}- \gamma^{2}\beta^{2}m^{2}(\xi_{A}-\xi_{B})^{2}  
\end{equation*}
\begin{equation}
\hspace{1.0cm}-p_{T,A}^{2}-p_{T,B}^{2}-2\: p_{T,A}\: p_{T,B} \:cos(\phi). 
\label{eq:b2} 
\end{equation}

The quantity $\phi$ in Eq. (\ref{eq:b2}) denotes the difference
in azimuthal angle of the two protons in the final state, 
$\phi = \phi_{A}\!-\!\phi_{B}$. The transverse momentum 
$p_{T}$ of the meson is given by 

\begin{equation}
p_{T}^{2} = p_{T,A}^{2} + p_{T,B}^{2} + 2\: p_{T,A}\: p_{T,B} \:cos(\phi). 
\label{eq:b3} 
\end{equation}

In the new coordinates $(u,v)$ defined in \mbox{Appendix A,}
the mass $M$ is expressed as

\begin{equation*}
M^{2} = \gamma^{2} \bigl[\beta(v_{A} + v_{B}) -
(u_{A} + u_{B}) + 2m\bigr]^{2}     
\end{equation*}
\begin{equation}
\hspace{1.cm} -\gamma^{2}\bigl[(v_{A}-v_{B})-\beta(u_{A}-u_{B})\bigr]^{2} - p_{T}^{2}.      
\label{eq:b4}
\end{equation}

It is convenient to introduce new variables $(u_{+},u_{-})$ and 
$(v_{+},v_{-})$ with a Jacobian determinant equal to unity,  

\begin{equation*}
u_{+} = \frac{u_{A}+u_{B}}{\sqrt{2}}, \hspace{.6cm}u_{-} = 
\frac{u_{A}-u_{B}}{\sqrt{2}}, 
\end{equation*}
\begin{equation}
v_{+} = \frac{v_{A}+v_{B}}{\sqrt{2}}, \hspace{.6cm}v_{-} = 
\frac{v_{A}-v_{B}}{\sqrt{2}}.
\label{eq:b5}
\end{equation}

The transverse momentum  $p_{T}$ is expressed in the variables
$(u_{+},u_{-},v_{+},v_{-},cos(\phi))$ as

\begin{equation}
p_{T}^{2}\!\!=\!F(u_{+},u_{-},v_{+},v_{-}\!)+G(u_{+},u_{-},v_{+},v_{-}\!)\:cos(\phi),
\label{eq:b6}
\end{equation}

with $F(u_{+},u_{-},v_{+},v_{-})$ and $G(u_{+},u_{-},v_{+},v_{-})$ defined as

\begin{equation}
F(u_{+},u_{-},v_{+},v_{-}) = u_{+}^{2} + u_{-}^{2} - v_{+}^{2} - v_{-}^{2} - 2 m^{2}, 
\label{eq:b7}
\end{equation}
\begin{equation*}
G(u_{+},u_{-},v_{+},v_{-})\!\!=\!\!\sqrt{(u_{+}\!\!+\!u_{-})^{2}\!-\!(v_{+}\!\!+\!v_{-})^{2}\!-\!\! 2m^{2}} 
\end{equation*}
\begin{equation}
\hspace{2.4cm}\times \sqrt{(u_{+}\!\!-\!u_{-})^{2}\!-\!(v_{+}\!\!-\!v_{-})^{2}\!-\!\!2m^{2}}. 
\label{eq:b8}
\end{equation}

The mass $M$ is expressed as

\begin{equation}
M^{2} = 2\gamma^{2} H(u_{+},u_{-},v_{+},v_{-}) - p_{T}^{2},  
\label{eq:b9}
\end{equation}

with $H(u_{+},u_{-},v_{+},v_{-})$ defined by

\begin{equation}
H(u_{+},u_{-},v_{+},v_{-})=(\beta v_{+}\!-\! u_{+}\!+\!\sqrt{2}m)^{2} -(\beta u_{-}\!-\!v_{-})^{2}.
\label{eq:b10}
\end{equation}

\section*{Appendix C}

\section*{Phase Space Parameterisation}

The parameterisation of the final state consisting of two
protons and a meson necessitates ten variables (not considering polarisation).
These ten parameters consist of the two three-momentum vectors of the protons,
in addition to the four-momentum of the central meson system. Imposing
conservation of energy-momentum of the initial state reduces the ten parameters
to six. These six variables can be taken as the squared four-momentum
transfer $t$, the fractional longitudinal momentum loss $\xi$ and the azimuthal
angle $\phi$ of each of the two protons. The distribution of Pomerons in the
proton used in this study does not have an azimuthal dependence, and the
azimuthal angle of one of the protons can hence be integrated out.
This integration leaves the difference in azimuthal angle of the two final
state protons as a phase space parameter. In the present study, this
azimuthal opening angle is uniformely distributed.
Our approach permits to assign a weight function to this opening angle,
thereby allowing the study of azimuthal angle correlations between
the two final state protons. The set of five variables
($u_{+},u_{-},v_{+},v_{-},cos(\phi)$) derived in Appendix B
defines the phase space of the final state unambiguously.

Differential cross sections can be derived by a kinematical transformation onto
the corresponding parameters. In particular, we would like to derive the meson
cross section differential with respect to meson \mbox{mass $M$} and meson
transverse momentum $p_{T}$, hence we are looking for a change of variables

\begin{equation}
(u_{+},u_{-},v_{+},v_{-},cos(\phi)) \longmapsto (u_{+},u_{-},v_{-},M,p_{T}),
\label{eq:c1}
\end{equation}

with an associated Jacobian determinant

\begin{equation*}
du_{+}du_{-}dv_{+}dv_{-}dcos(\phi) = 
\end{equation*}
\begin{equation*}
  \biggl|\frac{\partial(u_{+},u_{-},v_{+},v_{-},cos(\phi))}
       {\partial(u_{+},u_{-},v_{-},M,p_{T})}\biggr|\:du_{+}du_{-}dv_{-}dMdp_{T} =
\end{equation*}
\begin{equation}
  \biggl|\frac{\partial(v_{+},cos(\phi))}
       {\partial(M,p_{T})}\biggr|\:du_{+}du_{-}dv_{-}dMdp_{T}.
\label{eq:c2}
\end{equation}

The Jacobian determinant $J$ of Eq. \ref{eq:c2} can be evaluated with the use
of Eqs. (\ref{eq:b6}),(\ref{eq:b9}),


\begin{equation}
  J\!\!=\!\!\biggl|\frac{\partial(v_{+},cos(\phi))}{\partial(M,p_{T})}\biggr|\!\!=\!
  4 M p_{T} \biggl|\frac{\partial(M^{2},p_{T}^{2})}{\partial(v_{+},cos(\phi))}\biggr|^{-1}\!\!\!=\!\frac{2 M p_{T}}{G \gamma^{2} \frac{\partial H}{\partial v_{+}}}.
 \label{eq:c3}
\end{equation}

Here, $G$ and $H$ are the functions $G(u_{+},u_{-},v_{+},v_{-})$ and
$H(u_{+},u_{-},v_{+},v_{-})$ defined in Eqs. (\ref{eq:b8}),(\ref{eq:b10}). 

\vspace{0.6cm}



\end{document}